\documentclass{ws-ijmpa}
\usepackage[super,compress]{cite}
\usepackage{graphicx,subfigure}

\newcommand\beq{\begin{equation}}
\newcommand\eeq{\end{equation}}
\newcommand\beqn{\begin{eqnarray}}
\newcommand\eeqn{\end{eqnarray}}
\newcommand\nn{\nonumber}
\newcommand\fc{\frac}
\newcommand\lt{\left}
\newcommand\rt{\right}
\newcommand\pt{\partial}

\begin{document}
\markboth{Ke Yang, Yuan Zhong, Shao-Wen Wei and Yu-Xiao Liu}
{PURE GEOMETRIC BRANES AND MASS HIERARCHY}

%
\catchline{}{}{}{}{}
%

\title{PURE GEOMETRIC BRANES AND MASS HIERARCHY}

\author{Ke Yang\footnote{yangke09@lzu.edu.cn}$~^{a}$,
        Yuan Zhong\footnote{zhongy2009@lzu.edu.cn}$~^{a,b}$,
        Shao-Wen Wei\footnote{weishaow06@lzu.edu.cn}~$^a$,
        Yu-Xiao Liu\footnote{liuyx@lzu.edu.cn, corresponding author}~$^a$}

\address{$^a$Institute of Theoretical Physics, Lanzhou University, Lanzhou 730000, People's Republic of China \\
   {$^b$IFAE, Universitat Aut\`onoma de Barcelona, 08193 Bellaterra, Barcelona, Spain}
   }

\maketitle


\begin{abstract}
We consider a toy model with flat thin branes embedded in a 5-dimensional Weyl integrable manifold, where the geometric Weyl scalar provides the material that constitute the brane configurations. The brane configuration is similar to the Randall-Sundrum model. However, it is found that the massless graviton is localized on the brane with negative tension. So in order to solve the gauge hierarchy problem our world should be confined on the positive tension brane, and this is crucial to reproduce a correct Friedmann-like equation on the brane. The spacings of graviton mass spectrum are very tiny, but these massive gravitons are hidden in low energy experiments because they are weakly coupled with matter on the brane.

\end{abstract}

\ccode{PACS numbers: 04.50.-h, 11.27.+d}

\section{Introduction}

Motivated by string/M theory, the braneworld scenario has been intensively studied more than a decade (see Refs. \refcite{Rubakov2001,Kubyshin2001,Csaki2004,Dzhunushaliev2010} for an introduction). In this scenario, our universe is trapped in a four-dimensional submanifold (called brane) embedded in a fundamental multi-dimensional spacetime (called bulk). This scenario provides us a mechanism to solve some disturbing problems of high-energy physics, such as the gauge hierarchy problem (the problem of why the electroweak scale $M_{\text{EW}}\approx 1 $TeV is so different from the Planck scale $M_{\text{Pl}}\approx 10^{16} $TeV) and the cosmological constant problem\cite{Arkani-Hamed1998,Antoniadis1998,Randall1999,Randall1999a}. A landmark theory in braneworld scenario is the Randall-Sundrum (RS) model put forward in 1999 \cite{Randall1999,Randall1999a}. The configuration of RS1 model is two 3-branes locate at the boundaries of a 5-dimensional AdS slice \cite{Randall1999}. We are living on the brane with negative tension while massless spin-2 graviton is localized on the brane with positive tension. The hierarchy problem is solved by introducing a warp factor to warp the extra dimension, and further the warped extra dimension red-shifts the Planck scale to electroweak scale on the negative tension brane. When the radius of extra dimension approaches infinite, the negative tension brane is removed and one gets the RS2 model \cite{Randall1999a}. Even the extra dimension is noncompact now, effective 4-dimensional Newtonian gravity still can be recovered on the remaining brane. A lot of relevant issues have been widely discussed based on the braneworld models, see e.g., Refs. \refcite{Goldberger1999a,Lykken2000,DeWolfe2000,Gremm2000,Gremm2000a,Csaki2000,Campos2002,Wang2002,Charmousis2003,Bazeia2004a,Liu2007,Liu2008,Dzhunushaliev2009,Dzhunushaliev2010a,Liu2010b,Liu2011,Liu2012,Liu2013,Guo2014,Bazeia2014}.

In Refs. \refcite{Csaki1999,Cline1999,Shiromizu2000}, the researches show that confining our universe on the negative tension brane leads to a ¡°wrong-signed¡± Freedmann-like equation. This causes a severe cosmological problem in RS1 model. However, this problem can be avoided if our universe is confined in the positive tension brane. Inspired by the property of scalar-tensor theory that the graviton configuration is determined by both the warp factor and the background scalar, we generalize the RS1 model in scalar-tensor theory in the Ref. \refcite{Yang2012a}. In order to solve the gauge hierarchy problem in this model, our world should be confined on the positive tension brane rather than the negative tension one. Thus the severe cosmological problem is naturally overcome. Actually, since the scalar-tensor theory has a close relation with the Weyl geometry \cite{Romero2012a}, this toy model can be naturally embedded in a integrable Weyl geometry, and the geometric Weyl scalar provides the material to constitute the brane configurations. Therefore, we investigate this kind of generalized RS1 model in the integrable Weyl geometry in this paper.

The paper is organized as follows: In section \ref{Weyl_Geometry}, we simply introduce the Weyl geometry. In section \ref{The_Model}, we propose our model and solve the theory. In section \ref{Gravitational_Fluctuations}, gravitational fluctuations and some physical implications are discussed. Finally, brief conclusions are presented.

\section{Weyl Geometry}\label{Weyl_Geometry}

The prototype of Weyl geometry was proposed by Weyl in 1918 as a generalization of Riemann geometry to attempt to unify gravity with electromagnetism \cite{Weyl1918,ORaifeartaigh2000}. In Riemann geometry the covariant derivative is compatible with the metric, i.e., $\nabla_{K}g_{MN}=0$, while this is not hold in Weyl geometry and now the assumption is given by \cite{Salim1996,Dahia2007}
\beq
\nabla_{K}g_{MN}=\omega_{K}g_{MN}, \label{Weyl_Condition}
\eeq
where $\omega_{K}$ is a ``gauge" vector field on the Weyl manifold specified by the pair ($g_{MN},\omega_{K}$). Now consider an infinitesimal parallel transport $dx^{K}$, with the definition (\ref{Weyl_Condition}), the length $L=g_{MN}l^{M}l^{N}$ of a vector $l^{K}$ is changed by
\beq
dL=L\omega_{K}dx^{K}. \label{Length}
\eeq
This means Weyl geometry allows for possible variations in the length of vectors during parallel transport.

Nevertheless, Einstein firstly pointed out that the Weyl's original theory is inadequate as a physically acceptable one since the tick rates of atomic clocks would depend on their past history, known as the ``second clock effect". In order to overcome this defect, i.e., to hold the tick rate synchronization for clock traveling alone different paths between spacetime points $A$ and $B$, one just has to impose that circuit integral of Eq. (\ref{Length}) vanishes for an arbitrary closed path containing $A$ and $B$, i.e., $\oint dL=0$. With Stoke's theorem this condition implies that $\omega_{K}=\omega_{,K}$, namely the vector field is a gradient of a scalar denoted as $\omega$ here. This type of Weyl geometry is called a Weyl integrable manifold, specified as ($ g_{MN},\omega$). Since the scalar field $\omega$ specifies the Weyl manifold, it is actually a geometrical field. So with Eq. (\ref{Weyl_Condition}), the Weylian affine connection is expressed as
\beq
\Gamma_{MN}^P=\{ _{MN}^P \}-\frac{1}{2}\lt(\omega_{,M} \delta_N^P  + \omega_{,N}\delta _M^P-g_{MN}\omega ^{,P} \rt),
\eeq
where $\{_{MN}^P\}$ represents the Riemannian Christoffel symbol. Now with some algebra, the Weylian Ricci tensor $R_{MN}$ can be separated into a Riemannian Ricci tensor part $\hat R_{MN}$ plus some other terms with respect to the Weyl scalar (all the hatted magnitudes and operators are defined by the Christoffel symbol in what follows), i.e.,
\beq
R_{MN}=\hat R_{MN}+\fc{3}{2}\omega_{M;N}+\fc{1}{2}g_{MN}{\omega^P}_{;P}
+\fc{3}{4}\lt(\omega_{,M}\omega_{,N}-g_{MN}\omega_{,K}\omega^{,K}\rt),
\label{W_R_Ricci_Relation}
\eeq
where the semicolon is the covariant derivative $\hat\nabla$ with respect to the Christoffel symbol.

It is easy to check that the Weyl condition (\ref{Weyl_Condition}) is invariant under the Weyl rescaling,
\beqn
\bar g_{MN}=e^{-\phi}g_{MN},~~\bar\omega=\omega-\phi, \label{Weyl_Rescaling}
\eeqn
where $\phi$ is a smooth function defined on $M_W$. Especially, via a particular rescaling $\hat g_{MN}=e^{-\omega}g_{MN}$ and $\hat\omega=0$, namely, fixing $\phi=\omega$ in the Weyl rescaling, the condition (\ref{Weyl_Condition}) becomes $\nabla_{K}\hat{g}_{MN}=0$ and then the Weylian affine connection changes to Christoffel symbol
$\Gamma ^P_{MN}\rightarrow\{ _{MN}^P \}$. Thus the Riemann geometry specified by ($\hat g_{MN},0$) recovers in this case.

As a generalization of Riemann, Weyl geometry has been received more attentions on the studies of field theory and cosmology, see e.g., Refs. \refcite{Cheng1988,Perlick1991,Nishino2009,Moon2010}. Moreover, background scalar field with self-interaction potential are crucial for generating smooth thick brane configurations \cite{DeWolfe2000}, and a geometric scalar is naturally provided by Weyl integrable manifold, thus the authors in Refs. \refcite{Arias2002,Barbosa-Cendejas2005,Barbosa-Cendejas2006,Barbosa-Cendejas2008,Liu2010a} proposed some thick brane models based on the integrable Weyl geometry.

\section{The Model} \label{The_Model}

We start with a simple 5-dimensional pure geometric action on a Weyl integrable manifold $M^W_5$ with a kinetic term of Weyl scalar,
\beq
S_5^W=\frac{1}{2\kappa}\int_{M^W_5}{d^5x\sqrt{|g|}f(\omega)(R-\omega_{,K}\omega^{,K})},
\label{Weylian_Action_I}
\eeq
where $\kappa\equiv8\pi G_5$, which will be set to 1 for simplifying the notation, $G_5$ is the gravitational constant, and $f(\omega)$ is an arbitrary function of the Weyl scalar $\omega(z)$. In this
frame the Weylian Ricci tensor reads $R_{MN}=\Gamma_{MN,A}^A-\Gamma_{AM,N}^A +\Gamma _{MN}^P \Gamma_{PQ}^Q -
\Gamma_{MQ}^P \Gamma_{NP}^Q$.

So with the relation (\ref{W_R_Ricci_Relation}), the above Weylian action (\ref{Weylian_Action_I}) can be rewritten as
\beq
S^W_5=\fc{1}{2}\int_{M^W_5}{d^5x\sqrt{|g|}\lt[f(\omega)\hat R-\lt(4f(\omega)+4f_{\omega}(\omega)\rt)\omega_{,K}\omega^{,K}\rt]},\label{Weylian_Action_II}
\eeq
where $f_{\omega}(\omega)$ denotes the derivative with respect to the scalar $\omega$.  Then from the action (\ref{Weylian_Action_II}), the equations of motion are
\begin{subequations}\label{EOM}
\begin{eqnarray}
    f\hat G_{MN}&=&\lt(4f+4f_{\omega}\rt)\big(\omega_{,M}\omega_{,N}-\fc{1}{2}g_{MN}\omega^{,K}\omega_{,K}\big)
            +\big(\hat\nabla_M\hat\nabla_Nf-g_{MN}\hat\nabla^K\hat\nabla_K f \big),\label{EOM_1}\\
    f_{\omega} \hat R&=&-2\lt(4f+4f_{\omega}\rt)\hat\nabla^K\hat\nabla_K\omega-\lt(4f_{\omega}+4f_{\omega\omega}\rt)\omega^{,K}\omega_{,K}.\label{EOM_2}
\end{eqnarray}
\end{subequations}
Further, we can write the field equation (\ref{EOM_1}) in the form of Einstein equation by moving all Weyl scalar terms to the right-hand side to compose an effective stress-energy tensor. This approach has been proved to be useful in practice in scalar-tensor gravity and $f(R)$ gravity \cite{Sotiriou2010}, namely,
\beq
\hat G_{MN}=\hat T_{MN}, \label{Efective_EEq}
\eeq
where $\hat G_{MN}$ is the effective Einstein tensor and $\hat T_{MN}$ is the effective energy-momentum tensor given by
\beqn
\hat T_{MN}&=&\big(4+4\fc{f_{\omega}}{f}\big)\big[\omega_{,M}\omega_{,N}-\fc{1}{2}g_{MN}\omega^{,K}\omega_{,K}\big]
            +\fc{f_{\omega}}{f}\big[\hat\nabla_M\hat\nabla_N\omega-g_{MN}\hat\nabla^K\hat\nabla_K\omega\big]\nn\\
           &+&\fc{f_{\omega\omega}}{f}\big[\omega_{,M}\omega_{,N}-g_{MN}\omega^{,K}\omega_{,K}\big]. \label{Effective_EMT}
\eeqn
Furthermore, the effective energy density $\hat \rho$ referring to static observers with the 4-velocity $U^{M}$ is defined
as $\hat \rho=\hat T_{MN}U^{M}U^{N}=-T^0_0$.

When the kinetic term is absent, the Weyl action (\ref{Weylian_Action_I}) is invariant under the Weyl rescaling (\ref{Weyl_Rescaling}) if and only if $f(\omega)=e^{-\fc{3}{2}\omega}$. Thus for $f(\omega)=e^{-\fc{3}{2}\omega}$, the scalar and the metric cannot be totally fixed. In Ref. \refcite{Yang2012a}, we chose $f(\omega)=e^{k\omega}$ with $k\neq-\fc{3}{2}$ to break the invariance. However, here the additional scalar kinetic term breaks the invariance, and when this invariance is broken, the Weyl scalar field transforms into an physical observable matter degree of freedom to generate the brane configuration. Therefore, we still choose $f(\omega)=e^{k\omega}$ here but leave the parameter $k$ as an arbitrary constant.

We consider the embedment of 3-branes, which preserve 4-dimensional Poincar$\acute{\text{e}}$ invariance, in a 5-dimensional Weyl spacetime with an $S^1/Z_2$ orbifold extra dimension. The ansatz for the most general metric satisfying these properties is given by
\beq
ds^2_5=a^2(y)\eta_{\mu\nu} dx^\mu dx^\nu + dy^2, \label{NC_Metric}
\eeq
where $a(y)$ is the warp factor and $y\in[-y_b,y_b]$ denotes the physical coordinate of the extra dimension.
The orbifold is compact and its physical size is $[0,y_b]$. However, if the boundary $y_b\rightarrow\infty$, the topology of extra dimension $S^1/Z_2\rightarrow R^1/Z_2$. Therefore, the extra dimension will be noncompact anymore in this situation. After a coordinate transformation, $dy=a(y(z))dz$, one introduces a conformal metric which is useful for discussing the gravitational perturbations:
\beq
ds^2_5=a^2(z)(\eta_{\mu\nu} dx^\mu dx^\nu + dz^2), \label{C_Metric}
\eeq
here $z$ denotes the conformal coordinate of the extra dimension with $z\in[-z_b,z_b]$. This conformal coordinate still preserves $Z_2$-symmetry.

\begin{figure}[htb]
\begin{center}
\subfigure[$a(z)$]  {\label{warp_factor}
\includegraphics[width=4cm,height=4cm]{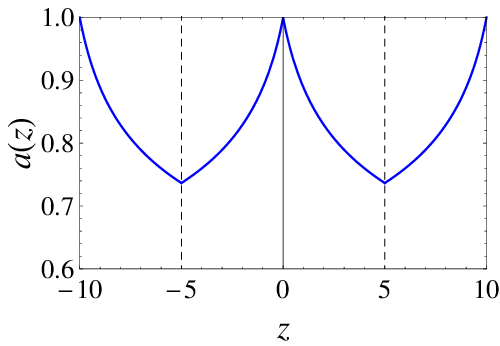}}
\hfill
\subfigure[$\omega(z)$]  {\label{scalar}
\includegraphics[width=4cm,height=4cm]{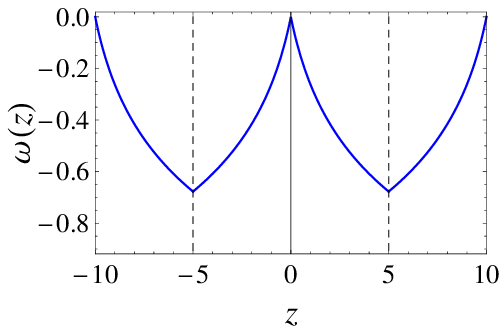}}
\hfill
\subfigure[$\hat\rho(z)$]  {\label{energy_density}
\includegraphics[width=4cm,height=4cm]{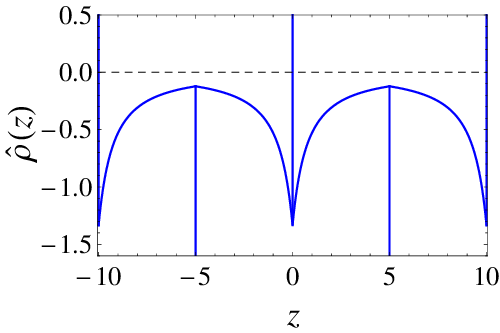}}
\end{center}
\caption{The shapes of the warp factor $a(z)$, scalar $\omega(z)$ and effective energy density $\hat\rho(z)$. The
parameters are set to be $\beta=1$, $k=-2$ and $z_b=5$.}
\label{Fig_Solution}
\end{figure}

In order to match the Poincar$\acute{\text{e}}$ invariance, $\omega(z)$ only refers to the extra dimension $z$. So the equations of motion (\ref{EOM}) can be simply expressed as
\begin{subequations}\label{Metric_EOM_2}
\begin{eqnarray}
    k\omega''+(k^2+2k+2)\omega'^2+2k\frac{a'}{a}\omega'+3\frac{a''}{a}&=&0, \label{Metric_EOM_2_1}\\
    (1+k)\omega'^2-2k\frac{a'}{a}\omega'-3\frac{a'^2}{a^2}&=&0,,\label{Metric_EOM_2_2}\\
    2(1+k)(\omega''+3\frac{a'}{a}\omega')+k(1+k)\omega'^2-k(\frac{a'^2}{a^2}+2\frac{a''}{a})&=&0.\label{Metric_EOM_2_3}
\end{eqnarray}
\end{subequations}
where the prime dentes the derivative with respect to $z$. From Eq. (\ref{Metric_EOM_2_2}), we easily get
\beqn
\omega'_{\pm}=\fc{k\pm\alpha}{1+k}\fc{a'}{a}, \label{So_In_EOM_2}
\eeqn
where $\alpha\equiv\sqrt{3+3k+k^2}$. After substituting $\omega'_\pm$ into Eq. (\ref{Metric_EOM_2_1}) or (\ref{Metric_EOM_2_3}), we achieve two branches of solution $(a_-, \omega_+)$ and $(a_+, \omega_-)$, i.e.,
\beqn
a(z)_{\mp}&=&\left( {1 + \beta z} \right)^{\frac{1}{3} (1\mp\frac{k}{\alpha })},\\
\omega(z)_{\pm}&=&\pm\frac{1}{\alpha}{\ln ( {1 +\beta z} )},
\eeqn
where we have set some of the integral parameters to fix $a(0)=1, \omega(0)=0$ and left a free parameter $\beta>0$. For the branch $a_-$, ${1- \frac{k}{\alpha}}$ is always positive for any $k$. However, for the other branch $a_+$, ${1+ \frac{k}{\alpha}}$ is positive for $k>-1$, vanishing for $k=-1$, and negative for $k<-1$. Now in order to make sure that the null signal takes an infinite amount of time to travel from $z_b$ to $z=0$ when $z_b\rightarrow\infty$, as suggested in RS model \cite{Randall1999,Randall1999a}, we use the solution $(a_+, \omega_-)$. Further, after imposing the $Z_2$-symmetric condition, we arrive at
\beqn
a(z)&=&\left( {1 + \beta |z|} \right)^{\frac{1}{3} (1+\frac{k}{\alpha})},\label{Sol_WP}\\
\omega(z)&=&-\frac{1}{\alpha}{\ln ( {1 +\beta |z|} )},\label{Sol_Sc}
\eeqn
where the parameter $k<-1$. The warp factor and scalar are plotted in Fig. \ref{warp_factor} and \ref{scalar}, respectively. Since $\omega'(z)$ is not continuous at boundaries, form the definition (\ref{Effective_EMT}), there are delta functions in the energy density at boundaries. Thus in this case the effective energy density is found to be
\beq
\hat\rho(z)=\frac{{\beta ^2}}{{3{\alpha ^2}}}\fc{( k^2+\alpha k+6k+6 )}{{( {1 + z\beta } )}^{ \frac{2}{3}( {4 + \frac{k}{\alpha }})}} - \frac{2k\beta [\delta(z)-\delta(z-z_b)]}{\alpha( 1 + \beta |z| )^{ \frac{5}{3}+ \frac{2k}{3\alpha }}}. \label{Energy_Density}
\eeq
As shown in Fig. \ref{energy_density}, there is a delta function at each boundary, and the coefficients of the delta functions actually play the role of brane tension. So this solution suggests a thin brane model, and the model presents a positive tension brane at $z=0$ and a negative tension brane at $z=z_b$. This brane configuration is similar to the RS1 model, however, as we will see in next section, the gravitational fluctuations are quite different from that in RS1 model, hence we call the brane located at the origin as a visible brane here, which is our world living on, while the other located at $z_b$ is an invisible brane.

\section{Gravitational Fluctuations} \label{Gravitational_Fluctuations}

From the effective Einstein equation (\ref{Efective_EEq}), we consider the metric fluctuations in our model. Since here we only care about the tensor fluctuations, which refer to spin-2 gravitons, we use the gauges $g_{\mu5}=0$ and $g_{55}=1$ to remove the vector and scalar fluctuations for simplicity \cite{Rubakov2001}. So the fluctuational metric is given by
\beq
ds^2=a^2(z)[(\eta_{\mu\nu}+h_{\mu\nu}(x,z))dx^{\mu}dx^{\nu}+dz^2], \label{Perturbational_Metic}
\eeq
where $a^2(z)h_{\mu\nu}$ represents tensor fluctuations in the background spacetime. From this fluctuation metric, the first order fluctuational Einstein equation (\ref{Efective_EEq}) is expressed as:\\
$\mu\nu$-component
\beqn
&&\fc{1}{2}\eta_{\mu\nu}h''
    +\lt(\fc{3}{2}\fc{a'}{a}+\fc{1}{2}\fc{f_\omega}{f}\omega'\rt)\eta_{\mu\nu}h'
    +\fc{1}{2}\eta_{\mu\nu}\Box^{(4)}h
    -\fc{1}{2}h_{\mu\nu}''\nn\\
    &&+\fc{1}{2}\lt(\pt_{\mu}\pt_{\rho}h^{\rho}_{\nu}+\pt_{\nu}\pt_{\rho}h^{\rho}_{\mu}-\pt_{\mu}\pt_{\nu}h\rt)
    -\fc{1}{2}\eta_{\mu\nu}\pt_{\rho}\pt_{\sigma}h^{\rho\sigma}
    -\lt(\fc{3}{2}\fc{a'}{a}+\fc{1}{2}\fc{f_\omega}{f}\omega'\rt)h_{\mu\nu}'  \nn\\
    &&+\lt[2\omega'^2+\fc{f_{\omega}}{f}\lt(\omega''+2\fc{a'}{a}\omega'+2\omega'^2\rt)
    +\fc{f_{\omega\omega}}{f}\omega'^2+3\fc{a''}{a}\rt]h_{\mu\nu}-\fc{1}{2}\Box^{(4)}h_{\mu\nu}=\nn\\
&&\fc{f_\omega}{f}\lt(\pt_\mu\pt_\nu\tilde{\omega}-\eta_{\mu\nu}\Box^{(4)}\tilde{\omega}\rt)
    -\fc{f_\omega}{f}\eta_{\mu\nu}\tilde{\omega}'' -\lt[4\omega'+\fc{f_\omega}{f}\lt(2\fc{a'}{a}+4\omega'\rt)+2\fc{f_{\omega\omega}}{f}\omega'\rt]\eta_{\mu\nu}\tilde{\omega}'\nn\\
   && -\lt[\lt(\fc{f_{\omega\omega}}{f}-\fc{f_\omega^2}{f^2}\rt)\lt(\omega''+2\fc{a'}{a}\omega'+2\omega'^2\rt)
+\lt(\fc{f_{\omega\omega\omega}}{f}-\fc{f_{\omega}f_{\omega\omega}}{f^2}\rt)\omega'^2\rt]\eta_{\mu\nu}\tilde{\omega},
\label{Fluctuation_EOM_1}
\eeqn
$\mu5$-component
\beqn
\fc{1}{2}\pt_{\rho}{h^{\rho}_{\mu}}'-\fc{1}{2}\pt_{\mu}h'
=\fc{f_\omega}{f}\tilde{\omega}_{,\mu}+\left[\lt(4+4\fc{f_\omega}{f}
+\fc{f_{\omega\omega}}{f}\rt)\omega'-\fc{a'}{a}\fc{f_\omega}{f}\right]\tilde\omega_{,\mu},
\label{Fluctuation_EOM_2}
\eeqn
$55$-component
\beqn
 &&\fc{1}{2}\Box^{(4)}h
 +\lt(\fc{3}{2}\fc{a'}{a}+\fc{1}{2}\fc{f_\omega}{f}\omega'\rt)h'
 -\fc{1}{2}\pt_{\rho}\pt_{\sigma}h^{\rho\sigma}=  \nn\\
 &&\lt(\fc{f_{\omega\omega}}{f}
 -\fc{f_\omega^2}{f^2}\rt)\lt(2\omega'^2-4\fc{a'}{a}\omega'\rt)\tilde\omega
 -\fc{f_\omega}{f}\Box^{(4)}\tilde\omega
+\lt[4\omega'+4\fc{f_\omega}{f}\lt(\omega'-\fc{a'}{a}\rt)\rt]\tilde\omega',
\label{Fluctuation_EOM_3}
\eeqn
where $\Box^{(4)}=\eta^{\mu\nu}\pt_{\mu}\pt_{\nu}$, $h^{\lambda}_{\mu}=\eta^{\lambda\nu}h_{\mu\nu}$, and $h=\eta^{\mu\nu}h_{\mu\nu}$. And the first order component of scalar field equation (\ref{EOM_2}) is given by
\beqn
  &&\fc{f_\omega}{f}h''
     -\lt[4\omega'+4\fc{f_\omega}{f}\lt(\omega'-\fc{a'}{a}\rt)\rt]h'
     +\fc{f_\omega}{f}\lt(\Box^{(4)}h-\pt_{\rho}\pt_{\sigma}h^{\rho\sigma}\rt)= \nn\\
  &&\lt(4\fc{f_{\omega\omega}}{f}
         +4\fc{f_{\omega\omega\omega}}{f}
      \rt)\omega'^2\tilde\omega
     +\lt(8\fc{f_\omega}{f}
     +8\fc{f_{\omega\omega}}{f}\rt)\lt(\omega''\tilde\omega+3\fc{a'}{a}\omega'\tilde\omega+\omega'\tilde{\omega}'\rt)\nn\\
  &&-4\fc{f_{\omega\omega}}{f}(2\fc{a''}{a}+\fc{a'^2}{a^2})\tilde\omega+\lt(8+8\fc{f_\omega}{f}\rt)\lt(\tilde{\omega}''+3\fc{a'}{a}\tilde{\omega}'+\Box^{(4)}\tilde{\omega}\rt),
\label{Fluctuation_EOM_4}
\eeqn
where $\tilde\omega$ represents the fluctuation of Weyl scalar $\omega$.

Further, we consider the transverse-traceless (TT) components of $h_{\mu\nu}$ with notation $\bar{h}_{\mu\nu}$, which satisfy the TT condition,
\beqn
\bar{h}=\pt_{\nu}\bar{h}_{\mu}^{\nu}=0.
\eeqn

\begin{figure}[htb]
\begin{center}
\subfigure[$\Psi_0(z)$]  {\label{zero_mode}
\includegraphics[width=5cm,height=4cm]{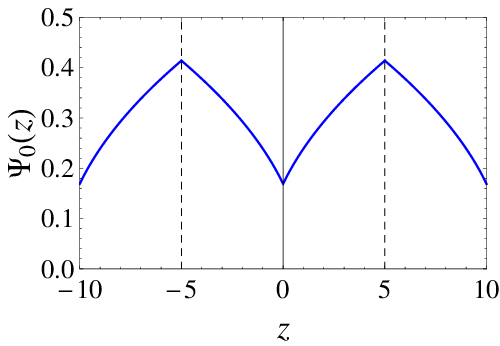}}
\hfill
\subfigure[$V(z)$]  {\label{potential}
\includegraphics[width=5cm,height=4cm]{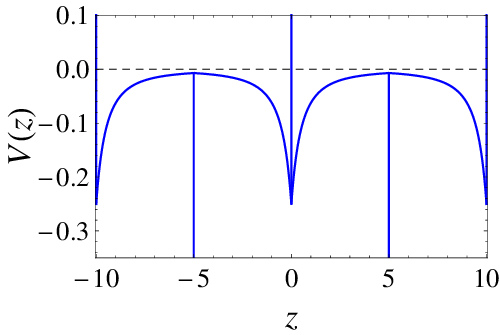}}
\end{center}
\caption{The shapes of the zero mode $\Psi_0(z)$ and effective potential $V(z)$. The
parameters are set to be $\beta=1$, and $z_b=5$.}
\label{Zero_mode_and_Potential}
\end{figure}

Inspection of above fluctuation equations reveals that Eqs. (\ref{Fluctuation_EOM_2}), (\ref{Fluctuation_EOM_3}) and (\ref{Fluctuation_EOM_4}) are purely non-TT components and all TT components are involved in Eq. (\ref{Fluctuation_EOM_1}).
Thus we make use of the TT projection operator for symmetric tensor field, namely, $P_{\mu\nu\sigma\rho}=\Pi_{\mu\sigma}\Pi_{\rho\nu}-\fc{1}{3}\Pi_{\mu\nu}\Pi_{\sigma\rho}$, to obtain the TT components of Eq. (\ref{Fluctuation_EOM_1}), where $\Pi_{\mu\nu}=\eta_{\mu\nu}-\fc{\pt_{\mu}\pt_{\nu}}{\Box^{(4)}}$. The TT projection operator has the properties that
${P_{\mu\nu}}^{\sigma\rho}h_{\sigma\rho}=\bar{h}_{\mu\nu}$, ${P_{\mu\nu}}^{\sigma\rho}\eta_{\sigma\rho}F=0$, and $
{P_{\mu\nu}}^{\sigma\rho}\pt_{\sigma}\pt_{\rho}F=0$, where $F$ is any scalar function. Furthermore, with the $\mu\nu$ component of Eq. (\ref{EOM_1}), we have a relation
\beq
-3\fc{a''}{a}=2\omega'^2+\fc{f_\omega}{f}\lt(\omega''+2\fc{a'}{a}+2\omega'^2\rt)+\fc{f_{\omega\omega}}{f}\omega'^2.
\eeq
Thus the TT components of equation (\ref{Fluctuation_EOM_1}) can be simplified as
\beq
\bar{h}''_{\mu\nu}+3\fc{a'}{a}\bar{h}_{\mu\nu}'+\fc{f_{\omega}}{f}\omega'\bar{h}'_{\mu\nu}+\Box^{(4)}{\bar{h}_{\mu\nu}}=0.\label{Fluctuation_Eq}
\eeq
Furthermore, we decompose $\bar{h}_{\mu\nu}$ in the form
\beq
\bar{h}_{\mu\nu}(x,z)=\varepsilon_{\mu\nu}(x)A^{-\fc{3}{2}}(z)\Psi(z)\label{Decomposition},
\eeq
where $A(z)\equiv a(z)f^{\fc{1}{3}}(\omega)$, and from the solution (\ref{Sol_WP}) and (\ref{Sol_Sc}), we simply arrive at $A(z)=(1+\beta |z|)^{1/3}$, and this is the same to that in Ref. \refcite{Yang2012a}. Then the four-dimensional mass $m$ of Kaluza-Klein (KK) excitations is defined by the Klein-Gordon equation $\Box^{(4)}{\varepsilon_{\mu\nu}(x)}=m^2 \varepsilon_{\mu\nu}(x)$. Finally,
a Schr$\ddot{\text{o}}$dinger-like equation can be obtained from Eq. (\ref{Fluctuation_Eq})
\beq
-\Psi''(z)+V(z)\Psi(z)=m^2\Psi(z),\label{Schrodinger_Eq}
\eeq
where the effective potential $V(z)$ is given by
\beq
V(z)=\fc{3}{2}\fc{A''}{A}+\fc{3}{4}\fc{A'^2}{A^2}= - \frac{\beta ^2}{4( {1 + \beta |z| } )^2}.\label{Effective_Potential}
\eeq
Note that the Hamiltonian in Eq. (\ref{Schrodinger_Eq}) can be factorized as the form
\beq
H=\lt(\fc{d}{dz}+\fc{3}{2}\fc{A'}{A}\rt)\lt(-\fc{d}{dz}+\fc{3}{2}\fc{A'}{A}\rt).
\eeq
Thus supersymmetric quantum mechanics ensures that there is no normalizable mode with $m^2<0$. There is no unstable tachyonic excitation in this model \cite{Csaki2000}. The spectrum of the eigenvalue in Eq. (\ref{Schrodinger_Eq}) gives the observed 4-dimensional graviton masses. Especially for the 4-dimensional massless graviton, by setting $m=0$ in Eq. (\ref{Schrodinger_Eq}), one easily gets the zero mode wave-function
\beq
\Psi_0(z)=\fc{A^{\fc{3}{2}}(z)}{N_0}=\fc{(1+\beta |z|)^{\fc{1}{2}}}{N_0},\label{Gravitational_zero_modes}
\eeq
where $N_0$ is the normalization constant. It shows that the scalar enters into the potential of Schr$\ddot{\text{o}}$dinger equation and the zero mode wave-function, which plays a crucial role to invert the massless graviton wave-function and make it localize on the brane at $z_b$ rather than on the one at origin as RS model, although the warp factor is still decrease towards to $z_b$. We plot the zero mode and effective potential in Fig. \ref{Zero_mode_and_Potential}.

Here we note that even these massless gravitational mode is normalizable in this compact extra dimension case, it is not normalizable anymore when $z_b\rightarrow\infty$. While this is capable in RS1 model and further ensure it removing the visible brane to transform to RS2 model \cite{Randall1999a}. Therefore, the necessity of compactifying the extra dimension is indeed crucial in our model. Furthermore, since the extra dimension is compact, all the massive KK modes are bound states with discrete eigenvalues.

Furthermore, the general solution of this equation is given in terms of Bessel functions
\beq
\Psi_n(z)=(1+\beta z)^{\fc{1}{2}}\lt[C_1\text{J}_{0}\big(m(z+1/\beta)\big)+C_2\text{Y}_{0}\big(m(z+1/\beta)\big)\rt],
\eeq
where $C_1$ and $C_2$ are $m$-dependent parameters. In addition, by imposing the Neumann boundary conditions $\pt_z \bar{h}_{\mu\nu}(x,z)=0$, we have
\beqn
\pt_{z}\Psi(z)|_{z=0}= \fc{\beta}{2}\Psi(z)|_{z=0},~~~
\pt_{z}\Psi(z)|_{z=z_b}=\fc{\beta}{2 (1 + \beta z_b)}\Psi(z)|_{z=z_b}.
\eeqn
So with the first boundary condition the wave function is
\beq
\Psi_n(z)=\fc{(1+\beta z)^{\fc{1}{2}}}{N_n}\lt[\text{J}_{0}\big(m(z+1/\beta)\big)+\alpha_n\text{Y}_{0}\big(m(z+1/\beta)\big)\rt],
\eeq
where $N_n$ is a normalization factor and $\alpha_n=-\text{J}_{1}(m/\beta)/\text{Y}_{1}(m/\beta)$. Then with the second boundary condition, we have the graviton mass spectrum which is satisfied
\beq
\fc{\text{J}_{1}\big(m_n(z_b+1/\beta)\big)}{\text{J}_{1}(m_n/\beta)}=\fc{\text{Y}_{1}\big(m_n(z_b+1/\beta)\big)}{\text{Y}_{1}(m_n/\beta)}.\label{Mass_Spectrum_Condition}
\eeq

Consider the the light modes in the long-range limit, i.e., $m_n/\beta\ll1$ and $1+\beta z_b\gg1$, then $\alpha_n\approx\fc{\pi m^2}{4\beta^2}\ll1$, thus from (\ref{Mass_Spectrum_Condition}), the spectrum is determined by $J_1(m_n(z_b+1/\beta))\approx0$, namely,
\beq
m_{n}=\fc{x_n}{z_b+1/\beta} \label{Mass_Spectrum},
\eeq
where $x_n$ satisfies $J_{1}(x_n)=0$, and $x_1=3.83$, $x_2=7.02$, $x_3=10.17,\cdots$. Thus the scale of mass splitting between the KK modes could be read off from this equation, as we could see in the following. Furthermore, from the normalization condition $\int^{z_b}_{-z_b}{\Psi_m\Psi_n}dz=\delta_{mn}$, the normalization factors are fixed by
\beq
N^2_0=2z_b+\beta z_b^2, ~~~N^2_n\approx 2z_b+\beta z_b^2,(n>0).
\eeq
Thus for low excited states, we have
\beq
\Psi_0(z)=\fc{(1+\beta z)^{\fc{1}{2}}}{\sqrt{2z_b+\beta z_b^2}},~~~\Psi_n(z)\approx \fc{(1+\beta z)^{\fc{1}{2}}}{\sqrt{2z_b+\beta z_b^2}}\text{J}_{0}\big(m_n(z+1/\beta)\big),(n>0). \label{Wave_Function_z}
\eeq
This implies that the lower KK states are also localized on the negative tension brane at the $z_b$. Further, with the decomposition (\ref{Decomposition}), the normalized wave-functions are given by
\beq
h^0_{\mu\nu}(x)=\fc{1}{\sqrt{2z_b+\beta z_b^2}}\varepsilon^{0}_{\mu\nu}(x), ~~~h^n_{\mu\nu}(x,z)\approx \fc{\text{J}_{0}\big(m_n(z+1/\beta)\big)}{\sqrt{2z_b+\beta z_b^2}}\varepsilon^{n}_{\mu\nu}(x),~(n>0). \label{Wave_Function}
\eeq

The interaction between gravitons and matter is achieved by including the action of SM matter fields $S_m=\int{d^5x\sqrt{g}L_m(x,z)}$. Varying this action with respect to the metric, one has $S_{\text{int}}=\fc{1}{2}\int{d^5x\sqrt{g}T^{MN}\delta{g}_{MN}}=\fc{\xi}{2}\int{d^5x}\sqrt{g}T^{MN}a^2h_{MN}$, where the factor $\xi=2/M_{*}^{3/2}$ is chosen to give the 5-dimensional field $h_{\mu\nu}$ a correct dimension, namely, $h_{\mu\nu}\rightarrow\xi h_{\mu\nu}$. When the matter fields located on the visible brane, i.e., $T^{MN}(x,z)=\tilde{T}^{\mu\nu}(x)\delta^{M}_{\mu}\delta^{N}_{\nu}\delta(z)$, where $\tilde{T}^{\mu\nu}(x)$ is the symmetric conserved Minkowski space energy-momentum tensor, it gives the usual form of the interaction Lagrangian in the 4-dimensional effective theory \cite{Davoudiasl2000}
\beq
L_{\text{int}}=\fc{\xi}{2}\tilde T^{\mu\nu}(x)h_{\mu\nu}(x,0). \label{Interaction_G_M}
\eeq
When $z=0$, $J_0(m_n/\beta)\approx1$, Eq. (\ref{Wave_Function_z}) shows $\Psi_n(x,0)\approx\Psi_0(x,0)$, thus (\ref{Wave_Function}) is reduced to $h^n_{\mu\nu}(x,0)=\varepsilon^n_{\mu\nu}(x)/(2z_b+\beta z_b^2),(n\geq0)$. Therefore Eq. (\ref{Interaction_G_M}) gives
\beq
L_{\text{int}}=\fc{\xi}{2\sqrt{2z_b+\beta z_b^2}}\tilde T^{\mu\nu}(x)\varepsilon^{n}_{\mu\nu}(x)=\tilde\xi\tilde T^{\mu\nu}(x)\varepsilon^{n}_{\mu\nu}(x),~~(n\geq0), \label{Coupling_Gravity_Matter}
\eeq
where $\tilde\xi=1/\big(M_{*}^{3/2}\sqrt{2z_b+\beta z_b^2}\big)$ is the effective coupling constant. It shows that the coupling of both massless graviton and massive KK gravitons to matters are of the same order in our model, while in RS1 model the couplings of massless and massive gravitons to the matter fields on its visible brane are quite different: the coupling of massless mode is of order $1/M_{\text{Pl}}$ and the coupling of massive KK modes is of order $1/\text{TeV}$.

As is well known, RS1 model can provide an exponential twist mechanism to solve the hierarchy problem relying on the exponential warp factor and two-brane configuration. Thus we simply discuss the possibility of solving the mass hierarchy problem based on our simple model. As in Ref. \refcite{Randall1999}, including only the massless zero mode in the fluctuation metric (\ref{Perturbational_Metic})
\beq
ds^2_5=a^2(z)\big[g^{(4)}_{\mu\nu}(x)dx^\mu dx^\nu + dz^2\big]=a^2(z)\lt[(\eta_{\mu\nu}+h^{0}_{\mu\nu}(x))dx^\mu dx^\nu + dz^2\rt]. \label{4D_Perturbational_NC_Metic}
\eeq
This will provide the gravitational fields in the effective theory. Here we note that since we have assumed that the Weyl scalar depends only on the extra dimension, the condition (\ref{Weyl_Condition}) on these thin branes is $\tilde\nabla_{\alpha}\tilde g_{\mu\nu}=\omega_{\alpha}\tilde g_{\mu\nu}=0$, and this implies that the connection ${\Gamma^{(4)\lambda}}_{\mu\nu}$ is just the Christoffel symbol constituted by the induced metric $\tilde g_{\mu\nu}(x)=a^2(z_0)g^{(4)}_{\mu\nu}(x)$ on the brane located at $z_0$. Thus the geometry of these thin branes is actually Riemannian.

Now calculating the contribution of the massless zero mode sector of the action (\ref{Weylian_Action_II}) gives us the 4-dimensional effective gravitational theory
\beqn
S^{W}_{5}\supset \fc{M_{*}^3}{2}\int_{M^W_5}{d^5x\sqrt{|g|}f(\omega)\hat R} \supset \fc{M_{*}^3}{2}\int_{-z_b}^{z_b}{dz a^3(z)f(\omega)}\int_{M^R_4}{d^4x\sqrt{|g^{(4)}|}\hat R^{(4)}},
\eeqn
where $M_{*}^{-3}=8\pi G_5$, and $\hat R^{(4)}$ is the four-dimensional Riemannian Ricci scalar made out of $g^{(4)}_{\mu \nu}=\eta_{\mu\nu}+h^{0}_{\mu\nu}(x)$. Thus the 4-dimensional effective scale of gravitational interaction is read from above equation as
\beq
M^2_{\text{Pl}}=M_{*}^3\int_{-z_b}^{z_b}{dz a^3(z)f(\omega)}=M^3_{*}\lt(2z_b+\beta z_b^2\rt), \label{Relation_M4_M5}
\eeq
where $M_{\text{Pl}}$ is our 4-dimensional Planck scale and $M^{-2}_{Pl}=8\pi G_N$ with $G_N$ the Newton's gravitational constant.

On the other hand, consider a fundamental Higgs field on the visible brane
\beq
S_{H}\supset\int{  d^4x\sqrt{|\tilde g|}\lt[-\tilde{g}^{\mu\nu}D_{\mu}H^{\dag}D_{\nu}H-\lambda(H^{\dag}H-v_0^2)^2 \rt]  },
\eeq
where $v_0$ is the vacuum expectation value (VEV) of Higgs scalar field and $\tilde g_{\mu\nu}(x)$ the induced metric on the visible brane. In case 2, the warp factor is $a(0)=1$, thus it leads to the induced metric $\tilde g_{\mu\nu}(x)=g^{(4)}_{\mu\nu}(x)$. Plugging this into the above action we obtain the effective action for the Higgs field
\beq
S_{H}\supset\int{d^4x \sqrt{|g^{(4)}|}\lt[-g^{(4)\mu\nu}D^{\mu}H^{\dag}D_{\nu}H-\lambda(H^{\dag}H-v_0^2)^2 \rt]}.
\eeq
Here it shows that the Higgs field is just the usually 4-dimensional canonically normalized form on the brane, thus the VEV scale takes its physical value. Furthermore, since the Higgs VEV sets all the mass parameters, any effective physical mass $m_{\text{vis}}$ on this visible brane is identical to its mass parameter $m_{*}$ in the fundamental theory
\beq
m_{\text{vis}}=m_{*}. \label{Mass_Hierarchy}
\eeq

Eqs. (\ref{Relation_M4_M5}) and (\ref{Mass_Hierarchy}) provide us a mechanism that can be used to solve the mass hierarchy problem. Thus if we set all the fundamental parameters $M_{*}, p, v_0$ to be of order of TeV scale in our theory, then from (\ref{Relation_M4_M5}), we only require $\beta z_b\approx10^{16}$ to provide a large twist of the two scale $\beta M^2_{\text{Pl}}\approx10^{32}M^3_{*}$.

On the other hand, in this case the mass spectrum given by Eq. (\ref{Mass_Spectrum}) is $m_{n}=\fc{x_n}{z_b+1/\beta}\approx(\beta z_b)^{-1}x_n\beta\approx10^{-4}$eV. Thus the spacing of the KK gravitons is quite small and it seems that the masses are tiny enough to allow energetics to produce these KK gravitons in colliders. Nevertheless, as is shown in Eq. (\ref{Wave_Function_z}), both the massless mode and lower massive KK modes are suppressed on the visible brane and localized on the other invisible one. Thus the effective coupling constant in (\ref{Coupling_Gravity_Matter}) is set to be $\tilde\xi=1/\big(M_{*}^{3/2}\sqrt{2z_b+\beta z_b^2}\big)=1/M_{\text{Pl}}$. It means that both massless and massive gravitons interact with our matter fields on the brane with 4-dimensional gravitational strength $1/M_{\text{Pl}}$, therefore these light KK gravitons can certainly not be seen individually. This is in contrast with that of RS1 model where the spacing of KK gravitons is of order of TeV scale, but similar to ADD model \cite{Arkani-Hamed1998,Antoniadis1998} with two extra dimensions, where the spacing is also tiny and about $10^{-3}$eV.

Even though each KK graviton weakly couples to the matter on the brane, the total number of KK modes increases with the physical energy scale. So following the same spirit of the large extra dimension model \cite{Arkani-Hamed1998,Antoniadis1998}, we roughly evaluate the cross section for real emission of these KK gravitons. Such as for the process $e^+ + e^- \rightarrow \gamma + \text{KK graviton}$, the total cross section is roughly given by
\beq
\sigma\sim\fc{\alpha}{M^{2}_{\text{Pl}}}N(E)\sim\fc{\alpha}{M^{2}_{\text{Pl}}}\fc{E}{10^{-4}\text{eV}}=10^{-16}\fc{\alpha E}{\text{TeV}^3},
\eeq
where $\alpha$ is the fine-structure constant, $E$ is the relevant physical energy scale, and $N(E)$ is the number of KK modes. Therefore, their collective contribution is also small enough to ensure that the model is not contrary to the experimental observations.

\section{Conclusions} \label{Conclusions}

In this paper, we have considered a toy brane model based on the Weyl integrable geometry to overcome the cosmological problem of RS1 model.
We achieve the thin brane solutions with two flat branes located at the boundaries.
For the Weyl scalar depending only on the extra dimension, the geometry on the brane is still Riemannian.

Since the massless graviton wave-function in gravitational fluctuations (\ref{Gravitational_zero_modes}) is determined by both the warp factor and the Weyl scalar, so it is very interesting that even the solution is totally different from the solutions in Ref. \refcite{Yang2012a} for the additional scalar kinetic term, but their gravitation perturbations are exactly the same because of their same Schr$\ddot{\text{o}}$dinger-like equation. Although the graviton mass spectra can not be distinguished between two models, their mass spectra of bulk matter fields are different, since the spectra relies on the form of warp factor \cite{Liu2013a}. As we have expected, the massless graviton is also localized on the negative tension brane at $z_b$, and this is crucial for moving our world to the positive tension brane to solve the cosmological problem. On the other hand, as gravitational zero modes are not normalizable when $z\rightarrow\infty$, compactifying the extra dimension is indeed necessary in our model.

In order to produce a large twist between the 4-dimensional Planck scale and 5-dimensional fundamental scale in the model, our world should locates on the positive tension brane at the origin, and this ensures the correct brane Freedmann-like equation. With all the fundamental parameters set to the order of TeV scale and $\beta z_b\approx 10^{16}$, a large enough hierarchy is generated between the 4-dimensional Planck scale and the fundamental electroweak scale. Moreover, the spacing of KK modes is found to be very tiny, namely, about the order of $10^{-4}$eV. It is more similar to ADD model with two extra dimensions. However, since the lower massive KK gravitons are suppressed on the visible brane, their interaction strength with the matter fields on the brane is just the weak 4-dimensional gravitation strength $1/M_{\text{Pl}}$. So the light KK gravitons cannot be seen individually in colliders.
With some roughly evaluation, we show that their collective contribution is also hidden in the low energy experimental observations.

\section*{Acknowledgments}

This work was supported by the National Natural Science Foundation of China (Grants No. 11205074 and No. 11375075), and the Fundamental Research Funds for the Central Universities (Grants No. lzujbky-2013-18).



\end{document}